\documentstyle[]{l-aa}

\begin{document}

\thesaurus{12 (12.03.1; 12.03.3; 12.03.4; 12.12.1)} 

\title{Cosmology with Galaxy Clusters}
\subtitle{II. Projection Effects on Hubble Constant and Gas Mass Fraction}
\author{Asantha R. Cooray}
\institute{Department of Astronomy and Astrophysics, University of Chicago, Chicago IL 60637, USA. E-mail: asante@hyde.uchicago.edu}
\date{Received:  May 20th 1998; accepted: August 18th 1998}
\maketitle

\begin{abstract} 

It is well known
that a combined analysis of 
the Sunyaev-Zel'dovich (SZ) effect and the X-ray emission observations 
can be used to determine the 
angular diameter distance to galaxy clusters, from which the Hubble
constant is derived. The present values of the Hubble constant derived
through the SZ/X-ray route have a broad distribution ranging from 30 to
70 km s$^{-1}$ Mpc$^{-1}$. We show that this
broad distribution is primarily due to the projection effect
of aspherical clusters which have been modeled
using spherical geometries. The projection effect is also
expected to broaden the measured gas mass fraction in galaxy
clusters. However, the projection effect 
either under- or overestimate the Hubble
constant and the gas mass fraction in an opposite manner,
producing an anticorrelation.
Using the published data for SZ/X-ray clusters, 
we show that the current Hubble constant distribution is negatively correlated
with the measured gas mass fraction for same clusters, suggesting
that the projection effects are present in current
results. If the gas mass fraction of galaxy clusters, when measured out
to an outer hydrostatic radius is constant, it may be possible to
account for the line of sight geometry of galaxy clusters.
However, to perform such an analysis, an independent measurement of
the total mass of galaxy clusters, such as from weak lensing, is
needed. Using the weak lensing, optical velocity dispersion, SZ and X-ray 
data, we outline an
alternative method to calculate the Hubble constant, which is subjected less
to projection effect than the present method based on only the SZ and X-ray
data. For A2163, the Hubble constant based on published SZ, X-ray and
weak lensing observations is 49 $\pm$ 29 km s$^{-1}$ Mpc$^{-1}$.

\end{abstract}



\section{Introduction}

Over the last few years, there has been a tremendous increase in 
the study of galaxy clusters 
as cosmological probes, initially through the
use of X-ray emission observations, and in recent years,
through the use of Sunyaev-Zel'dovich (SZ) effect.
Briefly, the SZ effect is
a distortion of the cosmic microwave background
(CMB) radiation by inverse-Compton scattering of 
thermal electrons within the hot intracluster medium
(Sunyaev \& Zel'dovich 1980; see Birkinshaw 1998 for a recent review). 
By combining the SZ intensity change and the X-ray emission observations,
the angular diameter distance, $D_{\rm A}$, 
to a cluster can be derived 
(e.g., Cavaliere {\it et al.} 1977).
Combining the distance measurement with redshift
allows a determination of the 
Hubble constant, H$_0$.
On the other hand,
angular diameter distances with redshift can be used to constrain
cosmological world models. 

The accuracy of the Hubble constant determined from a SZ and X-ray analysis
depends on the assumptions. 
Using numerical simulations, Inagaki et al. (1995) and 
Roettiger {\it et al.} (1997) showed that the
Hubble constant measured through the SZ effect can seriously be
affected by systematic effects, which include the assumption
of isothermality, cluster gas clumping, and asphericity.
The Hubble constant can also be affected
by statistical effects, including cluster peculiar velocities and
astrophysical confusions, such as radio sources \& CMB primary anisotropies.
The latter 
statistical effects are expected to produce a broad distribution in the 
Hubble constant measured for a sample of galaxy clusters, while the former
 systematic effects are expected to offset the Hubble constant
from the true value.

In recent years, several other effects have also been suggested
to explain the difference
between the SZ and X-ray Hubble constant and the ones derived from
other techniques. These include the
preferential removal of the lensed background radio sources
in SZ surveys (Loeb \& Refregier 1997), which would systematically
lower the Hubble constant by as much as 13\% for SZ observations
at 15 GHz, and gravitational lensing of the arcminute scale CMB anisotropy
(Cen 1998), which would broaden the 
Hubble constant distribution for a sample of galaxy clusters.
The first effect is in opposite direction to
the radio source contamination in SZ observations due to galaxy cluster
member radio sources, which dominate
the radio source number counts towards galaxy clusters.
As discussed in Cooray {\it et al.} (1998a), the two radio source effects
are likely to cancel out. The Loeb \& Refregier (1997) effect is also
not expected to occur for SZ observations at high frequencies.
The second effect, due to gravitational lensing of CMB anisotropy through
galaxy cluster potential, is not expected to be a dominant source of
error in the Hubble constant, given that Cen (1998) considered the
largest upper limits to arcminute scale anisotropies, which have not
yet been detected.

Apart from the SZ and X-ray Hubble constant, the gas mass fraction,
$f_{\rm gas}$, measurements
from X-ray (also using SZ, gravitational lensing  and
optical velocity dispersion measurements), can also 
be used to constrain the cosmological parameters.
The primary assumption in such an analysis is 
that the gas mass fraction, when measured out to
a standard (hydrostatic) radius is constant.
Evrard (1997) applied these arguments to a sample of galaxy clusters
using X-ray data, and put constraints on the cosmological
mass density of the universe, $\Omega_m$, with
some dependence on the Hubble constant.
Under the assumption that the cluster gas mass fraction is
 constant in a sample
of galaxy clusters, the apparent redshift evolution of the baryonic fraction
can also be used to constrain the cosmological parameters (e.g., Pen 1997). 
Cooray (1998) and Danos \& Pen (1998) used the present
X-ray gas mass fraction 
data to derive $\Omega_m < 0.6$ in a 
flat universe ($\Omega_m + \Omega_{\Lambda}
=1$) and $\Omega_m < 0.7$ in an open universe ($\Omega_{\Lambda}=0$; 
90\% C.I.). In Shimasaku (1997), the assumption of constant gas mass 
fraction in galaxy
clusters was used to put constrains on $\sigma_8$, the rms linear
fluctuations on scales of 8 h$^{-1}$ Mpc, and on $n$, the
slope of the fluctuation spectrum. 

Given the importance of SZ and X-ray emission
observations in cosmological studies, we initiated a program to study the
systematic effects in the present SZ and X-ray Hubble constant 
measurements and gas mass fraction measurements.
As part of this study, we found a negative 
correlation between the broad distribution of
the Hubble constant 
and the gas mass fraction measurements. We explain
this observation as due to 
a projection effect of aspherical clusters modeled with a
spherical geometry.
In Section 2, we present the effects of projection on
the Hubble constant and the gas mass fraction by projecting 
triaxial ellipsoidal clusters and extending the work of
Fabricant et al. (1984). The observational
 evidence for projection effects in the present Hubble constant
values based on SZ and X-ray route are presented in Section 3.
In section 4, we outline an alternative method to calculate the
Hubble constant, by combining SZ, X-ray, gravitational lensing, and
velocity dispersion measurements
of clusters, and which is subjected to less projection effects than
current method involving only the SZ and X-ray observations. 
We apply this technique to A2163 based on the published
observational data, and derive a new Hubble constant. A summary and
conclusions are presented in Section 5.

\section{Projection Effect of Aspherical Clusters}

In order to study the effect of aspherical clusters in present SZ and X-ray
Hubble constant, we extend the work of Fabricant et al. (1984) to
calculate the X-ray surface brightness and the SZ temperature change
produced by clusters with ellipsoidal geometries.
Independent of the cluster shape, 
the X-ray surface brightness towards a clusters is given by:
\begin{equation}
S_X = \frac{1}{4 \pi (1+z)^3} \int n ^{2}_{e} \Lambda_e dl,
\end{equation}
where $\Lambda_e \propto T_e^{1/2}$.
In order to model the electron number density profile within clusters, 
we consider the $\beta$-model, which can be written as:
\begin{equation}
n_{\rm e}(x_1,y_1,z_1) = n_{\rm e0} \left[1 + \frac{x_1^2+y_1^2}{r_1^2}+ \frac{z_1^2}{r_2^2}\right]^{-\frac{3 \beta}{2}},
\end{equation}
where $x_1,y_1$ and $z_1$ are coordinates of the ellipsoid axes, while
$r_1$ and $r_2$ are the observed semi-major and semi-minor axes. To
simplify the calculations, we assume that the symmetry axis $z_1$ 
is at an inclination angle $\theta$ to the line of sight along the observer,
which we take to be the
 $z$-axis. Following Fabricant et al. (1984, Appendix A), 
we integrate along the z-axis to
derive:
\begin{eqnarray}
S_X (x,y) = &\frac{\sqrt{\pi} n_{\rm e0}^2 \Lambda_{\rm e0}}{4 \pi (1+z)^3} \frac{\Gamma(3 \beta - \frac{1}{2})}{\Gamma(3 \beta)} \frac{r_1 r_2}{\sqrt{r_1^2 \cos^2{\theta} + r_2^2 \sin^2{\theta}}} \nonumber \\
&\times \left[1+ \frac{x^2}{r_1^2 \cos^2{\theta} + r_2^2 \sin^2{\theta}} + \frac{y^2}{r_1^2}\right]^{\frac{1}{2}-3 \beta}.
\end{eqnarray}

The other important observable towards clusters is the SZ effect, which is
given by:
\begin{equation}
\frac{\Delta T}{T_{\rm CMB}} = f(x)  
\int \left(\frac{k_B T_e}{m_e c^2}\right) n_e \sigma_T dl,
\end{equation}
where 
\begin{equation}
f(x) = \left[ \frac{x (e^{x}+1)}{e^{x}-1} -4 \right]
\end{equation}
is the frequency dependence with
$x = h \nu/k_B T_{\rm CMB}$, $T_{\rm CMB} = 2.728 \pm 0.002$ 
(Fixsen {\it et al.} 1994) and $\sigma_T$ is the 
cross section for Thomson scattering. 
The integral is performed along the line of sight through the
cluster. As with the X-ray surface brightness, we consider 
the same ellipsoidal shape to evaluate the observed SZ temperature change. 
Again by integrating along the line of sight, $z$-axis, we derive:
\begin{eqnarray}
\frac{\Delta T (x,y)}{T_{\rm CMB}} = & f(x) \sqrt{\pi} n_{\rm e0} T_{\rm e0} \frac{\Gamma(\frac{3 \beta}{2} - \frac{1}{2})}{\Gamma(\frac{3 \beta}{2})} \frac{r_1 r_2}{\sqrt{r_1^2 \cos^2{\theta} + r_2^2 \sin^2{\theta}}} \nonumber \\
&\times \left[1+ \frac{x^2}{r_1^2 \cos^2{\theta} + r_2^2 \sin^2{\theta}} + \frac{y^2}{r_1^2}\right]^{\frac{1}{2}-\frac{3 \beta}{2}}.  
\end{eqnarray}

The Hubble constant is usually derived by combining the X-ray brightness
and the SZ temperature change to eliminate the central number density 
$n_{\rm e0}$. By this combination, one can derive the observed
length of one of the axis, e.g.:
\begin{eqnarray}
r_2 = & \left [ \left( \frac{\Delta T_{\rm SZ} (x,y)^2}{S_{\rm X} (x,y)}\right) \left(\frac{m_{\rm e} c^2}{k_{\rm B} T_{\rm e0}}\right)^2 \frac{\Lambda_{\rm e0}}{4 \pi f(x)^2 T_{\rm CMB}^2 \sigma_{\rm T}^2 (1+z)^3} \right] \nonumber \\
& \times Z,
\end{eqnarray}
where $Z$ is the scale factor first introduced in Birkinshaw et al. (1991),
which can now be written as:
\begin{equation}
Z = \frac{\left(r_1^2 \cos^2{\theta} + r_2^2 \sin^2{\theta} \right)^{\frac{1}{2}}}{r_2}.
\end{equation}

When the symmetry axis of the cluster is along the line of sight 
($\theta=0$),
then $Z = r_1/r_2$ which is directly related to the observed
cluster ellipticity, while when the cluster is spherical ($r_1=r_2$),
$Z=1$, and no effects due to projection is present in the
data. In Eq. 7, we know from SZ and X-ray
observations all the quantities except the scale factor Z.
Therefore, the length of the cluster along the line of sight 
can be known up to a multiplicative
factor. The Hubble constant is derived based on the angular diameter
distance to the cluster, $D_A$, using an assumed cosmological
model, and the observed size of the axis, $\theta_{r_2} = r_2/D_A$, 
used to calculate the distance in Eq. 7.
The derived Hubble constant can be written as:
\begin{equation}
H_0 \propto \frac{\theta_{r_2}}{Z}.
\end{equation}
Based on observed ellipticities of galaxy clusters, we can estimate the
expected error in the Hubble constant. Using X-ray emission from
a sample of clusters, Mohr et al. (1995) showed that the median ellipticity
is $\sim$ 0.25. This suggest that the ratio 
$r_2/r_1$ is $\sim$ 0.7 if clusters are intrinsically prolate or
$\sim$ 1.5 if clusters oblate. Therefore, ignoring the effects due to 
inclination,
the Hubble constant as measured from SZ and X-ray observations of
an individual cluster can be offseted as much as 30\% to 50\%, based
on a spherical model of clusters where asphericity is ignored. 
Here, we have assumed that clusters are ellipsoids. The derived scale
factor in Eq.\ 8, as well as the numerical values, are likely to be
different if clusters are biaxial or triaxial. Recently, Zaroubi et al. (1998)
studied the projection effects of biaxial clusters and determined
$h \propto \sin\theta$, where $\theta$ is the inclination angle.
The observational evidence which suggest clusters are biaxial is limited.
For ellipsoidal clusters, we have determined that $h$ varies with both
the inclination angle and the sizes of semi-major and semi-minor axes.
For triaxial clusters, it is likely that $h$ will vary with
all three rotation angles and the length scales  of the three axes
that define the cluster. In a future paper, we plan to study the
projection effects of triaxial clusters; for the purpose of
this paper, we will only consider ellipsoids.

Apart from the Hubble constant, the projection effects are also present
in the total gas mass derived from the X-ray emission observations with
$M_{gas} \propto D_A^{5/2} Z^{1/2}$, and the total mass based on the
virial theorem using X-ray temperature as $M_{\rm total} \propto D_A Z^{-1}$.
Then, the gas mass fraction can be written as
 $f_{\rm gas} \propto D_{A}^{3/2} Z^{3/2}$.
Since $H_0 \propto Z^{-1}$ and the $f_{\rm gas} \propto Z^{3/2}$,
we expect the $H_0$ and $f_{\rm gas}$ to exhibit a negative
correlation, if both measurements are affected by the projection effect.

\section{Hubble Constant and Gas Fraction}

\begin{table*}[hbt]
\newlength{\digitwidth} \settowidth{\digitwidth}{\rm 0}
\catcode`?=\active \def?{\kern\digitwidth}
\caption{SZ Effect/X-ray $H_0$ Measurements and X-ray Gas Mass Fractions.}
\begin{tabular}{llclc}
\hline
Cluster & Redshift &$H_0$ (km s$^{-1}$ Mpc$^{-1}$)& $H_0$ Reference & $f_{\rm gas}$ ($r_{500}$) (h$_{50}^{-3/2}$) \\
\hline
A2256 & 0.0581 & 68$^{+21}_{-18}$ & Myers {\it et al.} 1997 & $0.11^{+0.03}_{-0.03}$\\
A478 & 0.0881 & 30$^{+17}_{-13}$ & Myers {\it et al.} 1997 & $0.25^{+0.03}_{-0.03}$\\
A2142 & 0.0899 & 46$^{+41}_{-28}$ & Myers {\it et al.} 1997 & 0.21$^{+0.01}_{-0.02}$ \\
A1413 & 0.143 & 44$^{+20}_{-15}$ & Saunders 1996 & 0.12$^{+0.01}_{-0.01}$\\
A2218 & 0.171 & 59 $\pm$ 23 & Birkinshaw \& Hughes 1994 & 0.16$^{+0.02}_{-0.02}$ \\
A2218 & 0.171 & 34$^{+18}_{-16}$ & Jones 1995 & 0.16$^{+0.02}_{-0.02}$\\
A665 & 0.182 & 46 $\pm$ 16 & Hughes \& Birkinshaw 1998b & 0.14$^{+0.02}_{-0.02}$ \\
A665 & 0.182 & 48$^{+19}_{-16}$ & Cooray {\it et al.} 1998c & 0.14$^{+0.02}_{-0.02}$\\
A2163 & 0.201 & 58$^{+39}_{-22}$ & Holzapfel {\it et al.} 1997 & 0.15$^{+0.01}_{-0.01}$\\
Cl0016+16 & 0.5455 & 47$^{+23}_{-15}$ & Hughes \& Birkinshaw 1998a & 0.17$^{+0.03}_{-0.03}$\\
\hline
\multicolumn{5}{@{}p{120mm}}{$H_0$ \& $f_{\rm gas}$ is calculated assuming
$\Omega_m=0.2$ and $\Omega_{\Lambda}=0$.}
\end{tabular}
\end{table*} 

Table 1 lists the Hubble constant values that have so far been
obtained from SZ observations
(Cooray {\it et al.} 1998b, see also Hughes 1997). 
These values have been calculated 
under the assumption of a spherical gas distribution
with a $\beta$ profile for the electron number density and an
isothermal atmosphere. 
For the same clusters, we compiled a list
of gas mass fraction measurements using X-ray, SZ, and gravitational lensing 
observations.
Most of the clusters in Table 1 have been analyzed
by Allen \& Fabian (1998), where they
included cooling flow corrections to the X-ray luminosity and the gas 
temperature. For the two clusters (A2256 \& Cl0016+16) 
for which $H_0$ measurements are available, but
not analyzed in Allen \& Fabian (1998),
we used the results from Buote \& Canizares (1996) 
and Neumann \& B\"ohringer (1996), respectively.
The gas mass fractions in Allen \& Fabian (1998) have been calculated
to a radius of 500 kpc, while for the A2256 and Cl0016+16, they
have been calculated to different radii, and also under different
cosmological models. Using the angular diameter distance dependence
on the gas mass fraction measurements with redshift (Cooray 1998), 
we converted all the gas mass fraction measurements to a cosmology of 
$\Omega_m=0.2$, $\Omega_{\Lambda}=0$, and
H$_0 = 50$ $h_{50}^{-1}$ km s$^{-1}$ Mpc$^{-1}$.
In order to facilitate comparison between the gas mass fractions
measured at various radii, we 
scaled them to the $r_{500}$ radius based on relations
presented by Evrard (1997).
The $r_{500}$ radius has been shown to be a good approximation to 
the outer hydrostatic boundary of
galaxy clusters (Evrard, Metzler, Navarro 1996).
We list the derived cluster gas mass fraction at the $r_{500}$ radius
in Table 1.

\begin{figure}[t]
\vspace{3in}
\caption[fig1]{The observed gas mass fraction of galaxy clusters 
and the SZ/X-ray Hubble constant. The vertical dashed line is the mean value of
the gas mass fraction. The solid line is the best-fit relation between
the $H_0$ values and the $f_{\rm gas}$ values, assuming
 $h \propto f_{\rm gas}^{-2/3}$. This line
is favored at $\sim$ 2 $\sigma$ confidence over a constant
$H_0$.}
\end{figure}

In Fig.\ 1, we show the calculated $f_{\rm gas}$ 
against $H_0$ values for each of the clusters. 
As shown, the gas fraction measurements have a broad distribution with
a scatter of $\sim$ 40\% from the mean value. 
A similar broadening of the Hubble constant,
from 30 to 70 km s$^{-1}$ Mpc$^{-1}$ with a mean of $\sim$ 50 km s$^{-1}$ 
Mpc$^{-1}$ is observed. The correlation is negative, and suggest that
clusters with high gas mass fraction measurements produces
Hubble constant values at the low end of the distribution, while the
opposite is seen for clusters with high gas mass fraction. 
The solid line in Fig.\ 1 is the best-fit relation between $h$ and
$f_{\rm gas}$ assuming $h \propto f_{\rm gas}^{-0.66}$.
 For values in Table 1, the best-fit line, when the 
slope between $h$ and $f_{\rm gas}$ is allowed to vary, scales as $h \propto
f_{\rm gas}^{-0.8 \pm 0.4}$, which is fully consistent with the
expected relation. Since the current SZ cluster sample is small,  
a careful study of a complete sample of galaxy clusters
are need to fully justify the projection effects between
SZ and X-ray derived Hubble constant and gas mass fractions values.
We derived a similar negative correlation between $h$ and $f_{\rm gas}$
when the cluster gas mass fraction is measured from SZ. For example,
Myers {\it et al.} (1997) derived a gas mass fraction of ($0.120 \pm 0.022$)
$h_{50}^{-1}$ for A2256, which is at the low end of the gas mass fraction 
values, while a gas mass fraction of ($0.33 \pm 0.028$) $h_{50}^{-1}$ was
derived for A478, which is the cluster at the high end. 
We note here that, as we discuss later, the SZ derived
gas mass fractions scale with $h$ as only $h^{-1}$, while X-ray derived
gas mass fractions, which are presented in Table 1, scale
with $h$ as $h^{-3/2}$. 
In comparison, the gas mass fractions derived from SZ and X-ray 
observations may be
 affected similar to the measurements based on only the X-ray data.
Additional probes of the total mass are the gravitational 
lensing measurements and the optical virial analysis of
internal galaxy velocity dispersion measurements. 
In the present SZ/X-ray sample, 
A2218 (Kneib {\it et al.} 1995) 
and A2163 (Squires {\it et al.} 1997) 
have lensing mass measurements. In both these clusters
total virial masses when measured using X-ray gas temperature,
 agrees with the weak
lensing mass measurements at large radii, and since these
two clusters are not the ones which are primarily responsible for
the observed negative correlation, we cannot state the effect of
lensing mass measurements on the above data. 
Also, in the present SZ cluster sample, 
A2256 and A2142 (Girardi {\it et al.} 1998), and
Cl0016+16 (Carlberg {\it et al.} 1997) have measured total masses
from optical virial analysis. These virial masses are in good agreement
with X-ray masses, allowing an independent robust measurement of
the total mass (Girardi {\it et al.} 1998).

Finally, there is a slight possibility that the observed broad distribution
and negative correlation
in $H_0$ and $f_{\rm gas}$ is not really present. The negative correlation
is only present at a level of $\sim$ 2 $\sigma$, assuming
that the errors in $h$ and $f_{\rm gas}$ are independent. The
removal of either one of the clusters at high or low end reduces the
negative correlation, decreasing the significance of the
observed correlation. 
However, both the Hubble constant and, possibly,
the gas mass fraction is expected to be constant, suggesting that a point, or
a region when considering errors in $H_0$ and $f_{\rm gas}$, 
is preferred. We rule out the possibility that both $H_0$ and $f_{\rm gas}$
are constants in the present data with a confidence greater than 95\%.

\subsection{Evidence for a Projection Effect?}

Usually, the broad distribution of the SZ and X-ray Hubble constants
has been explained in literature based on the 
expected systematic effects. 
The systematic effects in the gas mass fraction measurements are reviewed
in Evrard (1997) and Cooray (1998). We briefly discuss 
these systematic uncertainties in
the context of their combined effects on $H_0$ and $f_{\rm gas}$.

It has been suggested that cluster gas clumping may overestimate 
$H_0$ from the true value. As reviewed in Evrard (1997), cluster
gas clumping also overestimates $f_{\rm gas}$, suggesting
that if gas clumping is responsible for the observed trend, a positive
correlation should be present. The nonisothermality underestimates
$H_0$ by as much as 25\% 
(e.g. Roettiger {\it et al.} 1997). To
explain the distribution of $H_0$ values, the cluster
temperature profile from one cluster to another is expected
to be different.
However, Markevitch {\it et al.} (1997) showed the similarity between
temperature profiles of 30 clusters based on ASCA data (including
A478, A2142 \& A2256 in present sample). Since SZ and X-ray
structural fits weigh the gas distribution  differently, even
a similar temperature profile between clusters
 can be expected to cause the change
in the Hubble constant from one cluster to another.
Another result from the Markevitch {\it et al.} (1997)
study is that the $f_{\rm gas}$ measurements as measured using $\beta$-models
and standard isothermal assumption is underestimated. The
similarity of cluster temperature profiles also suggests that the gas
mass fractions are affected by changes in temperature from
one cluster to another. It is likely that the present
isothermal assumption has underestimated both $H_0$
 and $f_{\rm gas}$, and that temperature profiles
are responsible for the observed behavior. A large sample of clusters,
perhaps the same cluster sample studied by Markevitch {\it et al.} (1997),
should be studied in SZ to determine the exact effect of radial
temperature profiles on $H_0$, and its distribution.
 
The third possibility is the cluster asphericity. 
The effect of cluster projection on $H_0$ was first suggested
by Birkinshaw {\it et al.} (1991), who showed that the derived values for
$H_0$ can be offset by as much as a factor of 2
if the line of sight along the cluster is different by the same amount.
The present cluster isophotal ellipticities suggest that $H_0$
may be offset as much as $\pm$ 27\% (e.g., Holzapfel {\it et al.} 1997). 
The present $f_{\rm gas}$ distribution is suggestive of this behavior.
Cen (1997), using numerical simulations, studied the effects of
cluster projection on gas mass fraction measurements, and suggested
differences of the order $\sim$ 40\%. The $f_{\rm gas}$ distribution
is similar to what has been seen in Cen (1997).
It is more likely that the projection effects are causing the distribution of
$H_0$ and $f_{\rm gas}$ values, unless a systematic effect still not
seen in numerical simulations is physically present in galaxy clusters.
Such effects could come from effects due to variations in the 
temperature profiles from one cluster to another. For the rest of the discussion, we assume that the present values are affected by projection effects,
rather than temperature profiles.

\section{Hubble Constant without Projection Effects}

Here, we 
consider the possibility of deriving the Hubble constant in
a meaningful manner without any biases due to cluster projections. 
It has been suggested in literature that observations of a large
sample of galaxy clusters can be used to average out the dependence
on the scale factor $Z$ and to produce the true value of the Hubble constant,
which we define as $H_0^{\rm true}$ from individual Hubble constant
measurements, $H_0^{\rm i}$, in a large sample of clusters.
We investigate the possibility of such an averaging by considering
the different projections of clusters at different inclination angles.
Assuming the previously described ellipsoidal shape and
the effect of the scale factor $Z$ in the Hubble constant, 
we can over the all possible inclination angles $\theta$ and 
the ratio $r_2/r_1$ to derive
the expected average value of the Hubble constant $<H_0^{\rm i}>$:
\begin{equation}
<H_0^i> = \frac{1}{2}\left[ x + \frac{\sin^{-1}\sqrt{1-x^2}}{\sqrt{1-x^2}}\right] \times H_0^{\rm true},
\end{equation}
if all clusters are prolate, and
\begin{equation}
<H_0^i> = \frac{1}{2}\left[ x + \frac{\sinh^{-1}\sqrt{x^2-1}}{\sqrt{x^2-1}}\right] \times H_0^{\rm true},
\end{equation}
if all clusters are oblate.
Here $x = r_2/r_1$.
When all clusters are prolate and that the semi-major axis used to calculate
the Hubble constant, then the distribution has a mean of
$H_0^{\rm true}$. However, if the semi-minor axis is used, then
the average Hubble constant is underestimated from the true value
by about $\sim$ 10\%, assuming that the mean $r_2/r_1$ is 0.7 for
prolate clusters. If all clusters are oblate, and the semi-major
axis is used to derive the Hubble constant, then the mean of the distribution 
overestimates the true value of the Hubble constant by as much as
$\sim$ 20\%, if the mean $r_2/r_1$ is 1.5 for oblate clusters.
For oblate clusters, the true value of the Hubble constant can be obtained when
the semi-minor axis is used. However, in both oblate and prolate cases,
 the distribution has a large scatter requiring a large sample of galaxy
clusters to derive a reliable value of the Hubble constant.
A similar calculation can also be performed for the gas mass fraction
to estimate the nature of the value derived by averaging out a gas mass
fraction measurements for a large sample of clusters. Here again, a similar
offset as in the Hubble constant is present, and measurements
of gas mass fraction in a large sample of clusters are needed to put
reliable limits on the cosmological parameters, especially the mass
density of the universe based on cosmological baryon density (e.g.,
Evrard 1997).

So far, we have only considered the SZ and X-ray observations of
galaxy clusters. By combining weak lensing
observations towards galaxy clusters, we show  that
it may be possible to derive a reliable value of the Hubble constant
based on observations of a single cluster.
The gravitational lensing observations of galaxy clusters measure
the total mass along the line of sight through the cluster. The SZ effect
measures the gas mass along the line of sight, and thus, the ratio of
SZ gas mass to gravitational lensing total mass should yield a measurement
of the gas mass fraction independent of cluster shape assumptions
and asphericity. Here, we assume that
the cluster gas distribution exactly traces the cluster gravitational
potential due to dark matter, and that these two measurements are
affected equally by cluster shape. This is a reasonable assumption, but
however, it is likely that gas distribution does not follow
the dark matter potential, and that there may be some dependence
on the cluster shape between the two quantities. For now, assuming that
the gas mass fraction from SZ and gravitational lensing is not affected
by cluster projection, we outline a method to estimate the Hubble
constant independent of the scale factor $Z$.
The gas mass fraction based on SZ and lensing is
$f_{\rm SZ}^{\rm lens} \propto h^{-1}$, while the gas mass
fraction based on X-ray emission gas mass and the total mass based on
X-ray temperature is $f_{\rm X-ray}^{\rm temp} \propto h^{-3/2} Z^{3/2}$.
Since the two gas mass fraction measurements are expected to be the
same, then one can solve for a combination of $h$ and $Z$. However
to break the degeneracy between $h$ and $Z$ 
an additional observation or an assumption is needed.
In general, there are large number of clusters with X-ray measurements
and X-ray based gas mass fraction measurements. By averaging out the
gas mass fraction for such a large sample of
 clusters, we can estimate the universal
gas mass fraction value for clusters, e.g. $(0.060 \pm 0.002) h^{-3/2}$
(Evrard 1997; Cooray 1998). 
If assumed that this gas fraction is
valid for the cluster for which SZ and weak lensing observations are
available, we can then calculate the Hubble constant. 

We applied this 
to SZ, X-ray and weak lensing observations of galaxy cluster A2163.
The SZ observations of A2163 are presented in Holzapfel et al. (1997), while
weak lensing and X-ray observations are presented in Squires et al. (1996).
The SZ effect towards A2163 can be described with a $y$ ($\Delta T_{\rm SZ}/T_{\rm CMB}$) parameter of $3.07^{+0.54}_{-0.60} \times 10^{-4}$, which includes
various uncertainties described in Holzapfel et al. (1997). The weak lensing
observations of A2163 has been used to derive the total cluster mass
in Squires et al. (1996), and the lensing observations are most
sensitive out to a radius of $\sim$ 200$''$ (0.423 h$^{-1}$ Mpc) 
from the cluster center, where the total mass is
$(5 \pm 2) \times 10^{14}$ $h^{-1}$ $M_{\sun}$. 
Using the cluster model ($\beta$ and $r_c$) in
Holzapfel et al. (1997), we integrated the SZ temperature change to
this radius from cluster center along the line of sight to
derive a gas mass of $(4.3 \pm 2.5) \times 10^{13}$ $h^{-2}$ $M_{\sun}$.
This represents the gas mass within the cylindrical cut
across the cluster, and effectively probes the same region as the weak
lensing observations. The gas mass fraction based on the SZ effect and
the weak lensing total mass is $(0.086 \pm 0.060) h^{-1}$. 
When this gas mass is compared to the effective gas mass fraction of
clusters, $(0.060 \pm 0.020) h^{-3/2}$ (Evrard 1997; Cooray 1998), 
we obtain $h = 0.49 \pm 0.29$. We have slightly overestimated the error in the 
average gas mass fraction to take into account the fact that this
fraction is measured at the outer hydrostatic radius ($\sim$ 1 Mpc), 
and may not correspond to the value at the observed radius of A2163.
In Squires et al. (1997), the gas mass fraction was measured to be
$(0.07 \pm 0.03) h^{-3/2}$ for A2163, which is in agreement with
our universal value, but the value in Squires et al. (1997) may be
subjected to a scaling factor.
 The combined SZ/lensing gas mass fraction and the
average gas mass fraction for clusters result 
in a Hubble constant of $H_0 = 49
\pm 29$ km s$^{-1}$ Mpc$^{-1}$. Given that we used data from 2 different
papers in deriving this Hubble constant, it is likely that this
value may be subjected to unknown systematic effects between the two
studies. We strongly recommend that a careful analysis of cluster data be
carried out to derive the Hubble constant based on SZ, X-ray and
weak lensing observations. In addition, total
virial masses from velocity dispersion analysis should also
be considered in such an analysis to constrain the cluster shape.
It is likely that much stronger and
reliable result may be obtained through this method, instead of
just SZ and X-ray observations. In Holzapfel et al. (1997), the
Hubble constant was derived to be $\sim$ 60 km s$^{-1}$ Mpc$^{-1}$ for
an isothermal temperature model and $\sim$ 78 km s$^{-1}$ Mpc$^{-1}$ for
a hybrid temperature model. 
Our value is lower than these two values, but is in good agreement
with the average value of $H_0$ based on SZ and X-ray as tabulated in
Table 1, which is in agreement with the average gas mass fraction value.

\section{Conclusions}

Using the Hubble constant measurements based on SZ and X-ray, and the
gas mass fraction measurements, we have suggested
a possible systematic effect due to cluster projection. Even though,
cluster projection had been suggested as a possible systematic bias
in $H_0$ measurements, more attention has recently been given to
various {\it exotic} 
effects as a way to explain the broad distribution of
Hubble constant values. We have shown here the presence of
projection effects in the present $H_0$ and $f_{\rm gas}$ measurements and
have analytically calculated the effect of cluster projection in
deriving the Hubble constant. It is also assumed in literature
that for a large sample of clusters, the average  of the individual
Hubble constants, after making various corrections,
can be used to determine the true Hubble constant.
We have shown
here that this may not be easily possible,
and that when a random and large sample is
available with a mix of prolate and oblate clusters, 
the best that one could expect to obtain is 
a Hubble constant value within 10\% of the true value, unless
the distribution of ellipticities for cluster sample is 
carefully taken into account.
Thus, we strongly 
recommend that more attention be given to the cluster asphericity 
in deriving cosmologically important measurements 
such as Hubble constant and the 
cluster gas mass fraction.
For individual clusters, for which SZ observations are available, we have shown
that a combined study of SZ, X-ray, velocity dispersion measurements and 
weak lensing observations can be
used in a more physical manner to derive the Hubble constant.
Thus, we have demonstrated the usefulness of gravitational lensing
observations of galaxy clusters for cosmologically important studies,
and when combined, more meaningful results are expected to be produced
instead of just combining SZ and X-ray observations. We strongly
recommend that weak lensing observations and velocity
dispersion measurements be carried out to
test the reliability of Hubble constant values in Table 1, and to
complement SZ observations of clusters.

\acknowledgements
I would like to acknowledge useful discussions with John Carlstrom and
Bill Holzapfel. I would also like to thank the two referees, Mark Birkinshaw and an anonymous referee, for  detailed comments on the manuscript.
This study was partially supported by
the McCormick Fellowship at the University of Chicago,
and a Grant-In-Aid of Research from the National Academy of 
Sciences, awarded through Sigma Xi, the Scientific Research Society.

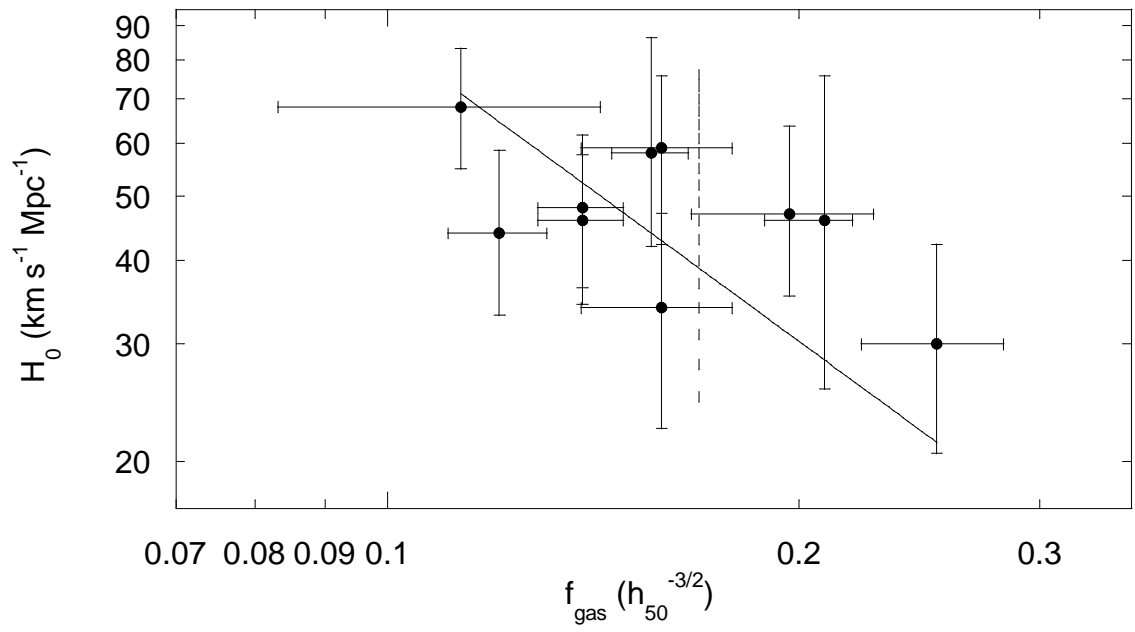
\end{document}